\documentclass[12pt]{article}
\usepackage{a4,epsf,overcite}
\usepackage[bf]{caption}
\makeatletter \renewcommand\@biblabel[1]{(#1)} \makeatother

\title{Effective interaction between star polymers} 
\author{A. Jusufi, M. Watzlawek, and H. L{\"o}wen\thanks{Also at: IFF, Forschungszentrum J{\"u}lich, D-52425 J{\"u}lich, Germany }  \\
       ~\hfill~ \\
       {\small Institut f{\"u}r Theoretische Physik II,}       \\ 
       {\small Heinrich-Heine-Universit{\"a}t D{\"u}sseldorf,}   \\
       {\small Universit{\"a}tsstra\ss e 1,}                   \\ 
       {\small D-40225 D{\"u}sseldorf, Germany}                \\
       ~\hfill~ \\
       }

\date{({\small {\em Macromolecules} {\bf 32}, 4470 (1999)})}

\begin{document}

\maketitle
\small\normalsize

\begin{abstract}
   The distance-resolved effective interaction between two star polymers 
   in a good solvent is calculated by Molecular Dynamics computer simulations. 
   The results are compared with a pair potential proposed recently 
   by Likos {\it et al.} [{\em Phys. Rev. Lett.} {\bf 1998}, {\em 80}, 4450] which
   is exponentially decaying for large distances and
   crosses over, at the corona diameter of the star,
   to an ultrasoft logarithmic repulsion for small distances. 
   Excellent agreement is found in a broad range of star arm numbers.~ \\

\end{abstract}

\small\normalsize

\newpage

Star polymers are hybrids between polymer-like entities 
and colloidal particles establishing an important link 
between these different systems, for recent reviews see ref
\citen{reviewSP,Gast}. The interpenetrability of two stars
is governed by the number of arms (or functionality) $f$, i.e. the
number of linear polymer chains attached to a central microscopic core.
For $f=1,2$ one recovers a system composed only of linear chains while
in the limit $f\to\infty$ one gets sterically-stabilized spherical
colloidal particles which behave like effective hard spheres 
\cite{PuseyLH,Hartmutr}.
Recent research \cite{Daoud,Grest,onestar,Willner} has mainly
focussed on polymer 
conformations of a {\it single} star. The only relevant
length scale of a single star is embodied in the  spatial 
extension of the monomers around the
core as given by the so-called corona diameter $\sigma$.

In order to predict macroscopic properties of a {\em concentrated} 
solution of {\em many} stars, 
one has, however, to proceed one step further: In any statistical
theory, the effective interaction between the stars is a necessary
input. This interaction, in general, comprises many-body terms.
For concentrations which are not too high, i.e. smaller than or
comparable to the overlap concentration $\rho^*\equiv 1/\sigma^3$,
triplet and higher-order terms are small and the system is dominated
by effective pairwise interactions. Recently, based on scaling theory 
\cite{Witten}, an explicit analytical expression for the effective
pair potential $V(r)$ was proposed in ref \citen{Likos}. 
This potential combines
a logarithmic form of the interaction potential for core-core separations $r$
smaller than $\sigma$ with an exponentially decaying interaction
of Yukawa-form for distances $r$ larger than $\sigma$:
\begin{equation} \label{potential}
   V (r) = \frac{5}{18} k_BT f^{3/2} 
           \cases{-\ln(\frac{r}{\sigma})+\frac{1}{1+\sqrt{f}/2} 
                   &; $r\leq\sigma$
           \cr
                   \frac{1}{1+\sqrt{f}/2}\left(\frac{\sigma}{r}\right)
                   \exp\left(-\frac{\sqrt{f}}{2\sigma}(r-\sigma)\right) 
                   &; $r>\sigma$
           \cr}
\end{equation}
Note that the potential strength simply scales with the thermal energy $k_BT$
since the repulsion between the stars is of purely entropic origin having 
a good solvent in mind. Both, the potential in eq \ref{potential}
and its associated force $F(r)=-dV(r)/dr$ are continuous at $r=\sigma$,
but $F(r)$ has an artificial cusp at $r=\sigma$.
The prefactor of the logarithm is fixed by scaling theory
\cite{Witten} while the
exponential decay length $2\sigma/\sqrt{f}$ 
is the diameter of the outermost blob within the
Daoud-Cotton model for one star polymer \cite{Daoud}.

For an arm number of $f=18$, this potential was tested against 
neutron scattering data and reasonable agreement was found
\cite{Likos}. Further experimental support comes from shear 
moduli measurements 
in the crystalline phase of many-arm-micelles \cite{Foerster}. 
Still, the scaling theory assumptions are strictly speaking only justified 
for core-core distances $r$ much smaller than $\sigma$, 
and the exponential decay length of the outermost blob size is an 
heuristic assumption. Hence the validity of the potential 
for arbitrary arm numbers can be questioned. In this paper, we test the 
pair potential against a microscopic model,
resolving the monomers of the chains, by extensive Molecular Dynamics 
computer simulations. To be specific, we use a simulation
model for star polymers developed by Grest et al. \cite{Grest}, 
which was applied in previous studies for single stars, 
and generalize it to a situation with two stars, which is
the minimal set-up to extract information about the effective 
interaction between two stars. 
The distance-resolved interaction force $F(r)$ is calculated for arm numbers
$f$ ranging from $f=5$ to $f=50$. Each arm contains $N$ monomers where
$N$ is varied from $50$ to $200$. As a result, we confirm the phenomenological
interaction potential in eq \ref{potential}; our simulation results are in perfect 
{\em quantitative} agreement with the theoretical prediction. This important result
enables a mapping of a star polymer solution onto a classical one-component
fluid \cite{HansenMcDonald} interacting via the effective ultra-soft 
pair potential of eq \ref{potential}, provided the star concentration 
does not exceed $\rho^*$.
This picture was anticipated in recent work, 
calculating the anomalous structure factor
of star polymer solutions \cite{our1} and the unusual phase diagram
including re-entrant melting \cite{Witten,our2} and anisotropic crystal 
structures \cite{our2}. So, our present work provides a theoretical 
justification of all these previous studies.

Let us first describe the simulation model \cite{Grest}: Each polymer arm consists
of $N$ effective monomers or ``beads" interacting via a purely
repulsive Lennard-Jones-like potential $V_{0}(r)$, 
where $r$ is the separation of the beads. $V_{0}(r)$ is obtained from
the usual Lennard-Jones potential $V_{LJ}(r)$ by cutting $V_{LJ}(r)$ 
at the position of the potential minimum $r_{m}=2^{1/6}\sigma_{LJ}$ and by shifting it by 
the constant value $V_{LJ}(r_{m})$ in order to obtain $V_{0}(r_{m})=0$: 
\begin{equation} \label{pot.lj}
   V_{0}(r)= \cases{4\epsilon\left[
                        \left(\frac{\sigma_{LJ}}{r}\right)^{12}
                        -\left(\frac{\sigma_{LJ}}{r}\right)^{6}
                        +\frac{1}{4}
                     \right] 
                   &; $r\leq 2^{1/6}\sigma_{LJ}$
             \cr
                   0 &; $r>2^{1/6}\sigma_{LJ}$
             \cr}
\end{equation}
Here, $\epsilon$ sets the energy scale and $\sigma_{LJ}$ the length scale
of the beads. The pure repulsion implies that we are dealing with
a good solvent. For neighbouring beads along the chains, the  
attractive FENE-potential \cite{Grest} $V_{ch}(r)$ is added to 
the interaction
\begin{equation} \label{pot.fene}
   V_{ch}= \cases{-15\epsilon\left(\frac{R_{0}}{\sigma_{LJ}}\right)^{2}
                   \ln\left[1-\left(\frac{r}{R_{0}}\right)^{2}\right]
                   &; $r\leq R_{0}$
             \cr
                   \infty &; $r>R_{0}$
             \cr}
\end{equation}
This interaction diverges at $r=R_0$, which determines the maximum 
relative displacement of two neighbouring beads. Henceforth, we fix
$R_0$ to be $1.5\sigma_{LJ}$. Then the total potential 
$V_{0}(r)+V_{ch}(r)$ between neighbouring monomers has a minimum
at $r\approx 0.97\sigma_{LJ}$. 
Furthermore, the central core particles of the two stars have a 
finite hard core radius $R_c$, and all monomers are interacting with 
the core particles via a modified repulsive interaction potential 
$V^c_{0}(r)$.  The introduction of a small hard core of the central
particles of the stars is necessary to accommodate the large number of
arms at small distances from the core \cite{Grest}.
Thus we take explicitly for the potential 
\begin{equation} \label{pot.core.lj}
   V^{c}_{0}(r)=\cases{ \infty &; $r\le R_{c}$
               \cr
                  V_{0}(r-R_{c}) &; $r>R_{c}$
               \cr} 
\end{equation}
In addition, the innermost monomers of each arm 
are interacting with their core via an attractive potential 
which is given by
\begin{equation} \label{pot.core.fene}
   V^{c}_{ch}(r)=\cases{ \infty &; $r\le R_{c}$
               \cr
                  V_{ch}(r-R_{c}) &; $r>R_{c}$
               \cr} 
\end{equation}
We note that exactly this simulation model was already used by Grest et al. in their
simulations of single star polymers \cite{Grest}. In our simulations,
the centers of the two stars are fixed at positions ${\vec R}_1$ and 
${\vec R}_2$ with a given distance $r=\left| {\vec R}_1 - {\vec R}_2 \right|$.
The total number of mobile monomers is $2fN$, which limits 
our studies to small $f$ and small $N$. In all simulations, 
the system is held at fixed temperature $T=1.2\epsilon/k_B$.
Under these circumstances, the effective force ${\vec F}_i$ 
acting on the $i$th star center is given as a canonical average
\begin{equation} \label{force}
   {\vec F}_{i} = \left\langle
      -{\vec \nabla}_{{\vec R}_{{i}}}\left(
          \sum_{k=1}^{2fN} V^{c}_{0}(|{\vec r_{k}}-{\vec R}_{i}|)
        + \sum_{l=1}^{f} V^{c}_{ch}(|{\vec r_{l}}-{\vec R}_{i}|)      
      \right)
   \right\rangle, 
\end{equation}
where in the first sum the repulsive interactions of the core with
{\em all} $2fN$ monomers in the system are considered, whereas the
second sum only accounts for the attractive interactions with the $f$
innermost monomers of the chains attached to the $i$th center. 
Obviously, due to symmetry, 
${\vec F}_1 = - {\vec F}_2$.
We use standard Molecular Dynamics simulations \cite{Allen} to equilibrate
the monomers and perform the statistical average 
$\langle ...\rangle$ over the monomers for the forces on the star centers. 
The timestep is typically
$\delta t = 0.002\tau$ (with $\tau
=\sqrt{m\sigma^{2}_{LJ}/\epsilon}$ being the associated time unit and
$m$ the monomer mass) and typically $120000$ steps are used for equilibration
and up to $t_{max}/\delta t=500000$ steps were simulated to gather statistics. It was carefully checked by 
monitoring the internal energy that the system had equilibrated.
A typical snapshot of two stars after equilibration is shown in 
Figure \ref{snapshot.plot}. As can be seen, the monomers of one star do not
penetrate much into the central region of the other star.

In order to check the code, we performed simulations of single stars
changing the arm numbers between $f=5$ and $f=50$ and the monomer numbers 
from $N=50$ to $N=200$. 
The corresponding results for the radius of gyration
$R_{G}^2 = \frac{1}{fN}\sum_{i=1}^{fN}({\vec r}_i - \vec{r}_{CM})^2$
(where $\vec{r}_{CM}$ is the center of mass of the whole star)
and the density profile of the monomers
are in very good agreement with the results given in 
ref \citen{Grest} and are well described by the scaling theory
of Daoud and Cotton \cite{Daoud}.  
For a detailed list of the simulation parameters and
the results for $R_{G}$, obtained from these single star simulations, 
see Table \ref{parameters.table}.

It should be noted that the effective forces on the star centers are the gradient of the
{\em effective} star-star potential. This effective potential, however,
differs in general from the monomer averaged potential energy of the star
centers \cite{Lowen}.
We therefore had to calculate the averaged forces ${\vec F}_i$ ($i=1,2$) from our
two star simulations to compare with the theoretical force as calculated from eq \ref{potential}.
In doing this, two difficulties are arising: i) The corona diameter
$\sigma$, which is the relevant length scale in the potential of eq
\ref{potential}, is not known {\em a priori}. ii) In contrast to the theory,
there is a finite core size $R_c$ in our simulation model.
 
As regards the first difficulty, $\sigma$ is usually
defined as the typical maximum range where a scaling behaviour of the 
monomer density around a single star center holds \cite{Daoud,Witten}. 
A statistical definition of $\sigma$, however, is missing. 
On the other hand, the radius of gyration $R_G$ has a clear definition 
as a canonical average, which can be calculated 
directly in simulations. We therefore use $R_G$, which was calculated
in the single star simulations, as basic length scale for our
simulation data and fit these data for $F=\left|{\vec F}_{i}\right|$ 
($i=1,2$) to the theoretical prediction for $F(r)$ 
using the least-square method and treating $\sigma$ as the single 
fit parameter. Afterwards we check how
the optimal value for $\sigma$ scales with $R_G$
as obtained from the single star simulations.
The procedure is consistent if the ratio $\lambda=\sigma/2R_G$ 
is independent of $f$. The second difficulty is resolved as follows:
A logarithmic potential for $r<\sigma$ implies that the data should 
fall on a straight line crossing the origin 
if one plots the {\it inverse force}, $1/F$, versus $r$ inside
the corona diameter. A typical plot is given in Figure \ref{invforce.plot}. 
In fact, the data fall on a straight line. Extrapolating the data, however, one 
does not hit the origin. The divergence of the force occurs already at
a finite distance $2R_{d}\approx 2R_c + \sigma_{LJ}$ which clearly has to be 
attributed to the presence of the finite core in the simulations. 
We note that both, $R_c$ and $R_d$, are microscopic length scales and of same
order of magnitude (see Table \ref{parameters.table}), thus being not
relevant for the macroscopic length $\sigma$ in the scaling regime.
We therefore normalize our distances by subtracting $2R_{d}$, thus 
matching the divergence of the force properly.
We emphasize that the slope of the straight line is in very good
agreement with the theoretical prediction, see again
Figure \ref{invforce.plot}. This implies that the theoretical 
prefactor ${5\over 18}f^{3/2}k_BT$ in eq \ref{potential} is confirmed by 
the computer simulations.
In Figure \ref{results.plot1}, 
%\ref{results.plot2}%
we show the effective force versus distance for five different
arm numbers $f$ and two monomer numbers $N$. 
The agreement with the theory is convincing for all $f$ and $N$. 
The consistency of our fitting procedure of the corona diameter
$\sigma$ is documented in Table \ref{parameters.table},
where the ratio $\lambda=\sigma/2R_{G}$ is given for different $f$ and $N$. 
We find $\lambda\approx 0.65$ independent of $f$. This value also coincides 
with the value used in ref \citen{Likos} to fit experimental data for 
$f=18$. We further note that $\lambda$ is independent of $N$, consistent
with scaling theory.
Finally, we prove the exponential decay of the force for 
distances larger than $\sigma$ by plotting the logarithm 
of the force versus distance in Figure \ref{lnforce.plot}
for one typical example. 
One clearly sees the crossover 
of the inner-core data to a straight line outside the core. The slope
is consistent with the theoretical one as determined by the outermost
blob size.

In conclusion, we have verified the ultra-soft pair 
interaction for star polymers by direct molecular simulations. It
is straightforward to generalize the method to two stars
confined in a periodically-repeated cubic cell in order to estimate
the shrinking of the corona diameter due to a finite star density.
Also, similar to charged colloids \cite{Hart},
triplets of stars should be considered to investigate the importance
of triplet interactions. Our future work lies along these directions.

~\hfill~ \\
We thank M. Schmidt and C. N. Likos for helpful remarks.\\
M. W. thanks the Deutsche Forschungsgemeinschaft for support
within SFB 237.
\pagebreak

\pagebreak
\section*{Tables}

\begin{table}[hbt]
\begin{center}
   \begin{tabular}{|c|c|c|c|c|c|c|c|} \hline
   $f$ & $N$ & $\delta t/\tau$ & $t_{max}/\delta t$ & $R_{G}/\sigma_{LJ}$ &
   $(R_{c}/\sigma_{LJ}) + 1/2$ & $R_{d}/\sigma_{LJ}$ & $\lambda$\\ \hline
   5 & 100 & 0.004 & 500000 & 13.53 & 0.65 & 1.39 & 0.61\\ 
   10 & 50 & 0.003 & 400000 & 10.37 & 1.1 & 1.21 & 0.66\\
   10 & 100 & 0.003 & 400000 & 16.18 & 1.1 & 0.89 & 0.64\\
   10 & 150 & 0.003 & 400000 & 19.71 & 1.1 & 1.31 & 0.60\\
   10 & 200 & 0.003 & 400000 & 24.52 & 1.1 & 1.42 & 0.67\\
   18 & 50 & 0.002 & 350000 & 11.19 & 1.25 & 1.38 & 0.68\\
   18 & 100 & 0.002 & 350000 & 17.10 & 1.25 & 1.64 & 0.65\\
   30 & 50 & 0.002 & 350000 & 12.22 & 1.6 & 1.89 & 0.66\\
   50 & 50 & 0.002 & 350000 & 13.35 & 1.8 & 2.40 & 0.69\\ \hline
   \end{tabular}
 \end{center}
   \caption{ \label{parameters.table} 
            List of the simulation parameters and the 
            corresponding results for
            $R_{G}$ and $\lambda=\sigma/2R_{G}$.            
           }
\end{table} 

%
%\pagebreak
\section*{Figure captions}
\begin{figure}[hbt]
%   \epsfxsize=10cm
%   \epsfysize=7.5cm
%   \hfill\epsfbox{./snapshot.ps}\hfill~
  \caption{ \label{snapshot.plot}
            Typical configuration for two stars with $f=10$ and
            $N=50$. The distance between the central core particles, which
            are shown as big black spheres, is $r=5.2\sigma_{LJ}$.   
            The gray and light gray monomers are belonging to the
            first and second star respectively.         
           }
\end{figure} 
\begin{figure}[hbt]
%   \epsfxsize=10cm
%   \epsfysize=7.5cm
%   \hfill\epsfbox{./invforce.ps}\hfill~
   \caption{ \label{invforce.plot} 
            Reduced inverse force $k_{B}T/(FR_{G})$
            between the centers of two star polymers
            (for $f=10$ and $N=50$) versus reduced distance $r/R_{G}$.
            The error bars were obtained by
            averaging over the results of $8$ independent
            simulations.           
           }
\end{figure} 
\begin{figure}[hbt]
%   \epsfxsize=10cm
%   \epsfysize=7.5cm
%   \hfill\epsfbox{./results1.ps}\hfill~
   \caption{ \label{results.plot1} 
            Simulation results (symbols) and theoretical results (lines) 
            for the reduced effective force $FR_{G}/k_{B}T$ versus
            reduced distance $(r-2R_{d})/R_{G}$.
            a) for $f=5, 10, 18$ and $N=100$,
            b) for $f=18, 30, 50$ and $N=50$.               
           }
\end{figure} 
\begin{figure}[hbt]
%   \epsfxsize=10cm
%   \epsfysize=7.5cm
%   \hfill\epsfbox{./lnforce.ps}\hfill~
   \caption{ \label{lnforce.plot} 
            Logarithm of the reduced force $\ln\left(FR_{G}/k_{B}T\right)$
            versus reduced distance $(r-2R_{d})/R_{G}$ for $f=10$ and $N=50$. The error bars were obtained by
            averaging over the results of $8$ independent
            simulations.            
           }
\end{figure} 
\vfill

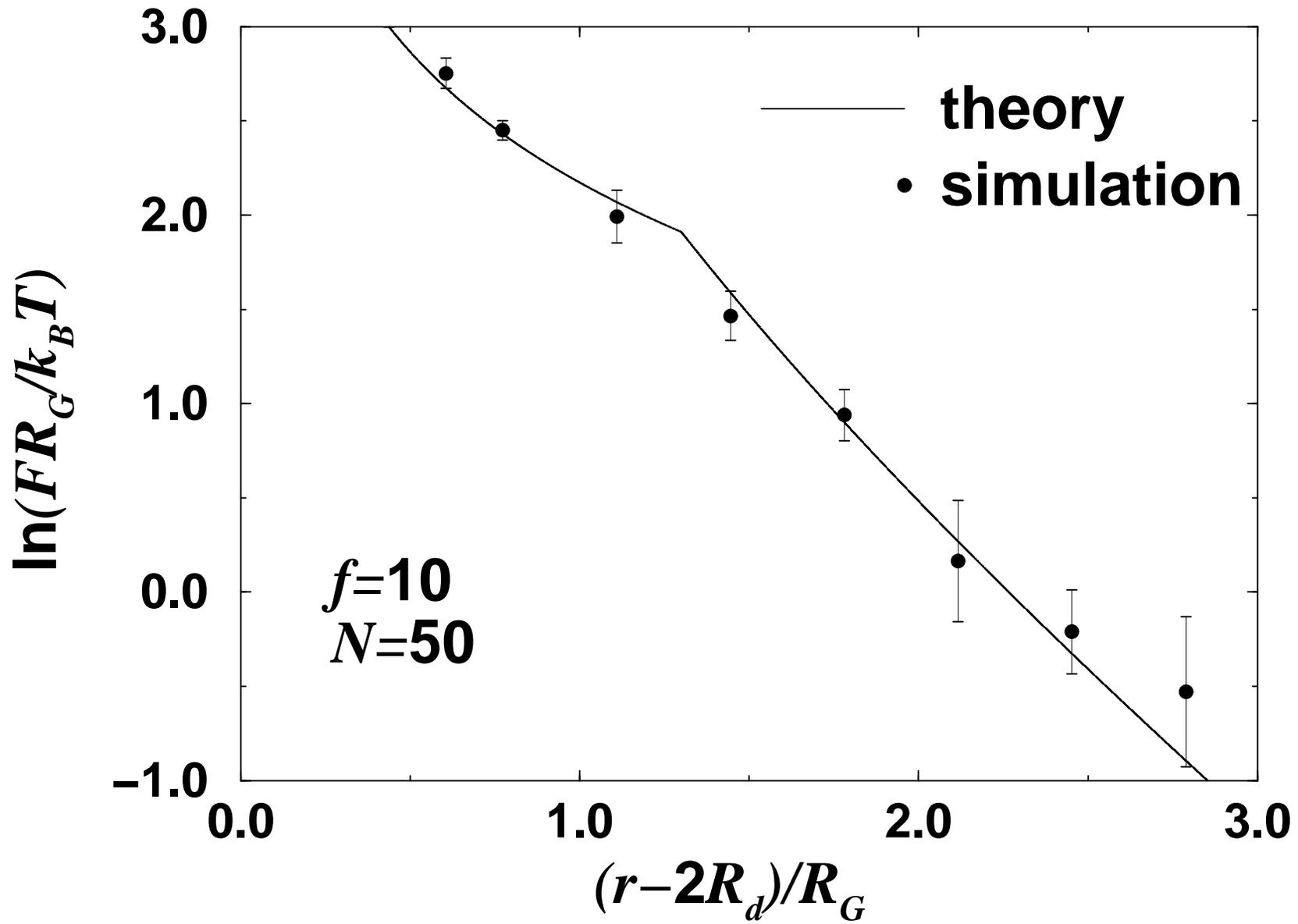
\end{document}